\documentclass[sigconf]{acmart}
\usepackage{multirow}
\usepackage{array}
\usepackage{color}
\usepackage{fancyhdr}
\usepackage{hyperref}
\def\BibTeX{{\rm B\kern-.05em{\sc i\kern-.025em b}\kern-.08emT\kern-.1667em\lower.7ex\hbox{E}\kern-.125emX}}

\copyrightyear{2019}
\acmYear{2019}
\acmConference[MM '19]{Proceedings of the 27th ACM International Conference on Multimedia}{October 21--25, 2019}{Nice, France}
\acmBooktitle{Proceedings of the 27th ACM International Conference on Multimedia (MM'19), October 21--25, 2019, Nice, France}
\acmPrice{15.00}
\acmDOI{10.1145/3343031.3350983}
\acmISBN{978-1-4503-6889-6/19/10}

\begin{document}
\fancyhead{}
%
\title{Progressive Retinex: Mutually Reinforced Illumination-Noise \\ Perception Network for Low Light Image Enhancement}
%


\author{Yang Wang$^{1}$,Yang Cao$^{1}$,Zheng-Jun Zha$^{1}$}
\authornote{Y. Wang and Y. Cao are co-first authors. Z.J. Zha is the corresponding author.}
\author{Jing Zhang$^{3}$,Zhiwei Xiong$^{1}$,Wei Zhang$^{2}$,Feng Wu$^{1}$}


\affiliation{%
  \institution{ywang120@mail.ustc.edu.cn \qquad {forrest, zhazj, zwxiong, fengwu}@ustc.edu.cn}
  jing.zhang1@sydney.edu.au \qquad davidzhang@sdu.edu.cn
}

\affiliation{%
  \institution{$^{1}$University of Science and Technology of China, Hefei, China \qquad $^{2}$Shandong University, Jinan, China}
  $^{3}$UBTECH Sydney Artificial Intelligence Centre, The University of Sydney, Sydney, Australia
}

\renewcommand{\shortauthors}{Anon.}

%
\begin{abstract}
Contrast enhancement and noise removal are coupled problems for low-light image enhancement. The existing Retinex based methods do not take the coupling relation into consideration, resulting in under or over-smoothing of the enhanced images. To address this issue, this paper presents a novel progressive Retinex framework, in which illumination and noise of low-light image are perceived in a mutually reinforced manner, leading to noise reduction low-light enhancement results. Specifically, two fully pointwise convolutional neural networks are devised to model the statistical regularities of ambient light and image noise respectively, and to leverage them as constraints to facilitate the mutual learning process. The proposed method not only suppresses the interference caused by the ambiguity between tiny textures and image noises, but also greatly improves the computational efficiency. Moreover, to solve the problem of insufficient training data, we propose an image synthesis strategy based on camera imaging model, which generates color images corrupted by illumination-dependent noises. Experimental results on both synthetic and real low-light images demonstrate the superiority of our proposed approaches against the State-Of-The-Art (SOTA) low-light enhancement methods.
\end{abstract}

%
%

%

%

%
\maketitle

\begin{figure}[t]
\begin{center}
\includegraphics[width=0.9\linewidth]{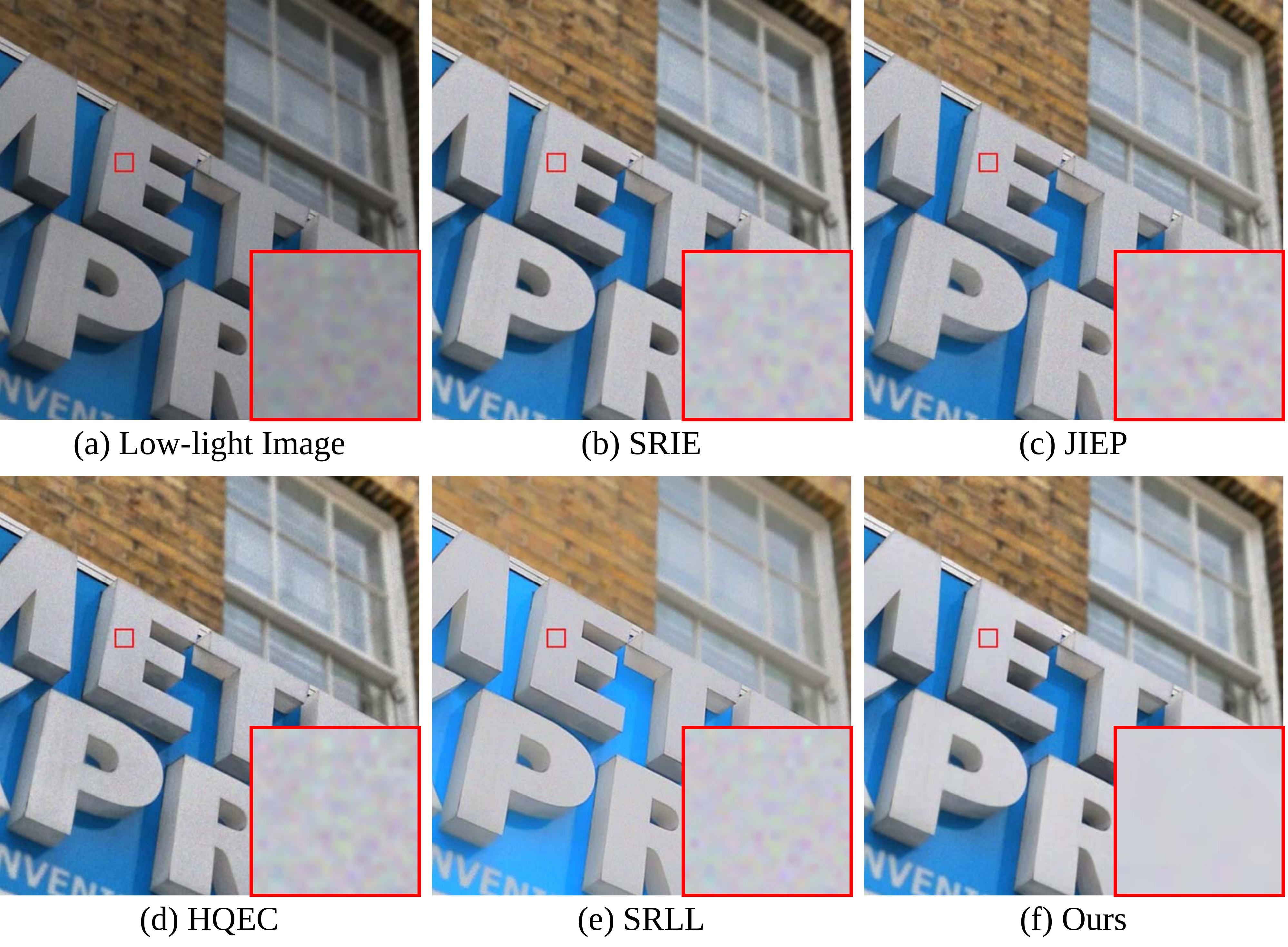}
\end{center}
   \caption{(a) Input low-light image; (b)-(f) enhancement result of SOTA methods (SRIE \cite{fu2016weighted}, JIEP \cite{cai2017joint}, HQEC \cite{zhang2018high}, SRLL  \cite{li2018structure}) and our method. As can be seen, there are still noise residuals in the enhanced images of existing methods, since the interaction between noise and illumination is neglected.}
\label{fig:fig1-ResultIllustration}
\end{figure}

\label{subsec:CouplingRelation}
\begin{figure*}
\begin{center}
\includegraphics[width=0.945\linewidth]{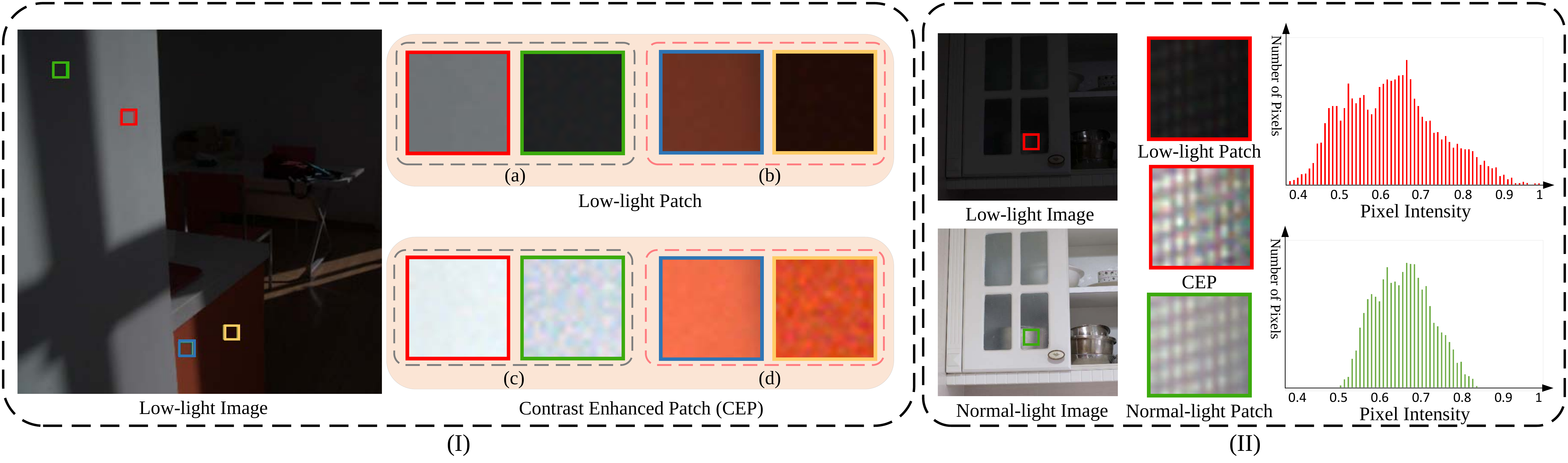}
\end{center}
   \caption{Illustration of coupling relation between illumination and noise. (a-b) The patch pairs with same image content and different illumination. (c-d) The contrast-enhanced results of (a-b) by using LIME \cite{guo2017lime}. As can be seen, the noise level is dependent on illumination. The histogram in \uppercase\expandafter{\romannumeral2} represents the intensity distribution of contrast-enhanced patch and normal-light patch respectively. Compared with the normal-light patch, the statistical distribution of CEP is greatly affected by the noise.}
\label{fig:fig9-CouplingRelation_2}
\end{figure*}

{\fontsize{8pt}{8pt}\selectfont
\textbf{ACM Reference Format:}\\
YangWang, Yang Cao, Zheng-Jun Zha,  Jing Zhang, Zhiwei Xiong, Wei Zhang and Feng Wu. 2019. Progressive Retinex: Mutually Reinforced Illumination-Noise Perception Network for LowLight Image Enhancement. In \textit{Proceedings of the 27th ACM Int'l Conference on Multimedia (MM'19), Oct. 21--25, 2019, Nice, France.} ACM, New York, NY, USA, \\9 pages.  https://doi.org/10.1145/3343031.3350983}

\section{Introduction}

Insufficient illumination in image capturing can significantly degrade the quality of images from many aspects, such as low visibility, contrast degradation, and high-level noise. These degradations not only cause unpleasant visual perception, but also hurt the performance of many computer vision systems which are designed for normal-light images. The Retinex that is motivated by human visual system (HVS) is an effective low-light image enhancement algorithm, providing color constancy and dynamic range compression. It assumes that observed images can be decomposed into the reflectance and illumination, denoted as $I(x,y) = R(x,y) \cdot L(x,y)$, where $R(x,y)$ is the reflectance determined by the characteristics of objects, and $L(x,y)$ is the illumination at each pixel $(x,y)$ which depends on the ambient light.

Retinex decomposition is known to be a mathematically ill-posed problem. One feasible solution for Retinex is to make assumptions of  the statistics of illumination and thus to devise specialized regularities for minimization. For example, some early heuristic Retinex algorithms work by assuming some regularities in the colors of natural objects viewed under canonical illumination, \emph{e.g.}, Multi-scale Retinex (MSR) \cite{jobson1997multiscale} and multi-scale Retinex with color restoration (MSRCR) \cite{rahman2004retinex}. Another practicable approach is to learn the statistical regularities, which formulates regression models of how the pixels related to illumination are distributed. Thus, the regression models indeed reveal the statistics of the pixels, and results in more general representations (\cite{wei2018deep} \cite{lore2017llnet}). Our method also falls into the latter category.

As shown in Fig.~\ref{fig:fig1-ResultIllustration}(a), image noise inevitably exists in low-light image, due to dark current and electronics shot in camera imaging \cite{liu2008automatic}\cite{tsin2001statistical}. To this end, M. Li \emph{et al}. introduces a noise term into the classic Retinex model to better formulate images captured under low-light conditions \cite{li2018structure}. Mathematically,
\begin{equation}
\begin{array}{l}
I = R \circ L + n ,
\end{array}
\label{0}
\end{equation}
Eq.(\ref{0}) implies that Retinex based low-light image enhancement indeed includes two tasks: contrast enhancement (determined by illumination map) and noise suppression (determined by noise level). The existing methods usually treat it as two separate tasks and solve them successively. For example, Joint-bilateral filter is applied to suppress the noise after the enhancement \cite{zhang2012enhancement}. Guo \emph{et al}. \cite{guo2017lime} attempts to further improve the visual quality by a post-processing via BM3D \cite{dabov2007image}. However, as shown in Fig.~\ref{fig:fig9-CouplingRelation_2}, there are coupling relations between the two tasks, which manifest as: 1) the noise level is indeed dependent on the intensity of illumination, 2) the existence of noise significantly affects the statistical distribution of illumination. Ignoring the coupling relationship may lead to inaccurate estimation of illumination or noise, resulting in under- or over-smoothing enhancement results (Fig.~\ref{fig:fig1-ResultIllustration}(b-e)).

To address this problem, this paper presents a novel progressive Retinex framework, in which illumination and noise of low-light image are perceived in a mutually reinforced manner to achieve Retinex-based image enhancement. Specifically, the statistical regularities of ambient light and image noise are modeled by two fully pointwise convolutional neural networks, and leveraged as constraints to facilitate the mutual learning process. The two modeling processes are progressively performed until obtaining stable results, and each of the preceding modeling processes benefits from the gradual improvement results in the other. The final estimations of illumination map and noise level are then leveraged for Retinex based image enhancement. Our proposed method not only suppresses the ambiguity between tiny textures and image noises, but also greatly improves the computational efficiency. Moreover, to solve the problem of insufficient training data, we propose a new data generation strategy based on camera imaging model. Experimental results on both synthetic and real low-light images demonstrate the superiority of our proposed method against the SOTA methods.

The contributions of this work can be summarized as:

$\bullet$ We present a novel progressive Retinex framework, in which illumination and noise of low-light image are perceived in a mutual reinforced manner, leading to visual pleasing enhanced results.

$\bullet$ We devise two fully pointwise convolutional neural networks to model the statistical regularities of ambient light and image noise respectively, and leverage them as constraints to facilitate the mutually learning process.

$\bullet$ Experimental results on synthetic and real data both demonstrate the superiority of our method over the existing Retinex methods and the other SOTA low-light enhancement approaches.

\label{subsec:Motivation-1}
\begin{figure*}
\begin{center}
\includegraphics[width=1\linewidth]{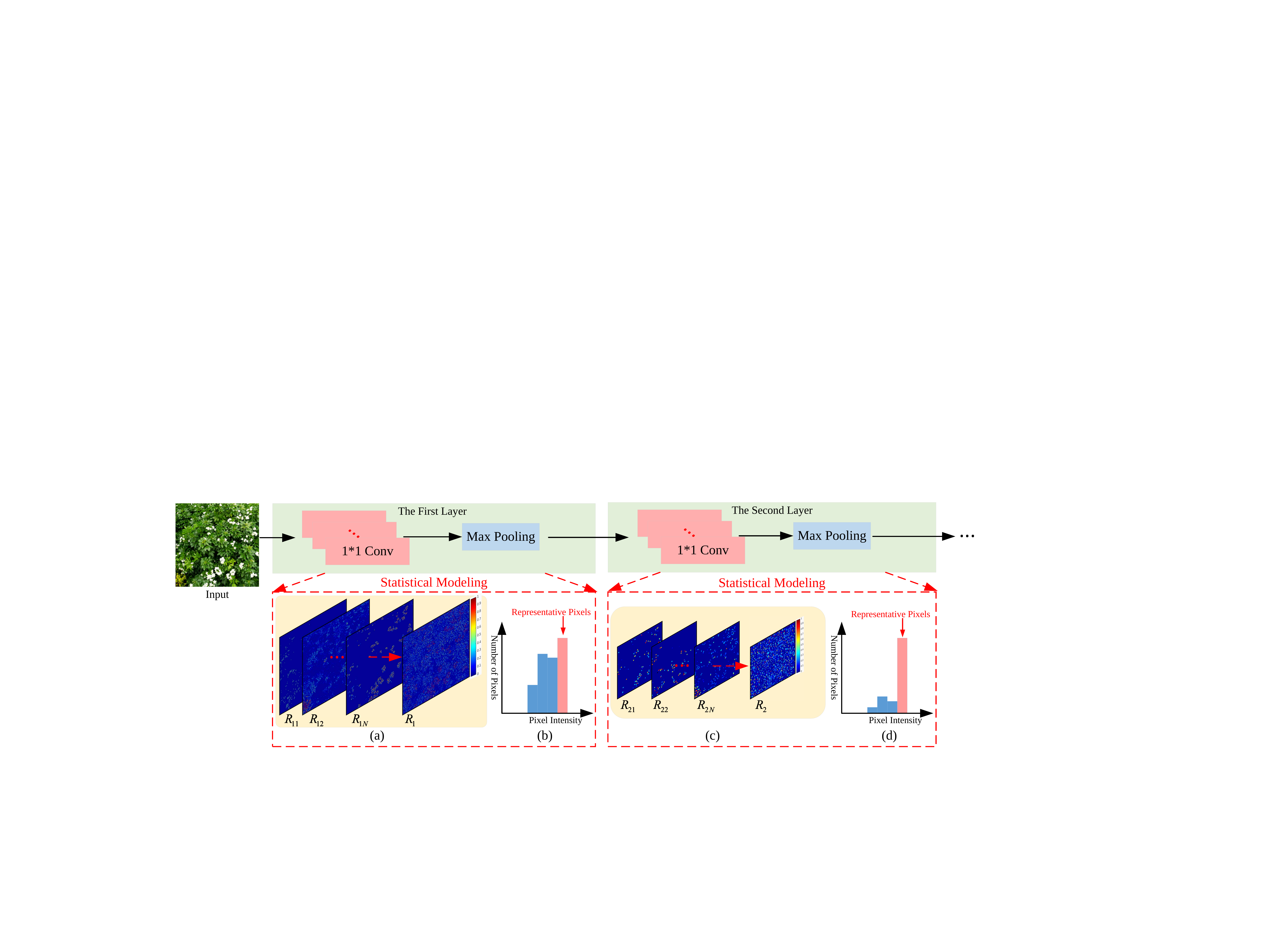}
\end{center}
\caption{Illustration of the pointwise convolution for statistical modeling in illumination estimation. The coordinates of the selected pixels is shown in (a). The statistical distribution of these pixels is shown in (b).  With network going deeper, the selected representative pixels can be further refined as shown in (c). And the low-confidence pixels will be discarded as shown in (d). We obtain ${R_{ij}}(i = 1,\,2;\;j = 1,\,2,\, \cdots \,,\,N)$ by projecting the feature map in pooling layer onto input. ${R_{ij}}(i = 1,\,2;\;j = 1,\,2,\, \cdots \,,\,N)$ denotes the coordinates of the selected pixels by each pointwise kernel ${K_{ij}}(i = 1,\,2;\;j = 1,\,2,\, \cdots \,,\,N)$. ${R_{i}}(i = 1,2)$ represents the coordinates of all selected pixels on each layer, \emph{i.e.}, the union set of ${R_{ij}}$. Warm color represents high confidence.}
\label{fig:Motivation-1}
\end{figure*}

\section{Related Work}
Extensive researches have been conducted to enhance the low-light images, which can be mainly divided into two categories: model-based methods and learning-based methods.

Model-based methods usually employ the model in Eq.(\ref{0}) or its variants to decompose an image into reflectance and illumination. According to Eq.(\ref{0}), the recovery of $R$ and $L$ from $I$ is an ill-posed inverse problem. A typical solution is to introduce a series of priors/assumptions of illumination and reflectance. In the PDE-based algorithm \cite{morel2010pde}, this ill-posed decomposition problem is modeled as a poisson problem by assuming that the reflectance changes at sharp edges and the illumination varies smoothly. MSR \cite{jobson1997multiscale} and MSRCR \cite{rahman2004retinex} incorporate illumination smoothness assumption and multi-scale information to get a robustness illumination estimation. In addition, The statistics of neighborhoods are used in \cite{forsyth1988novel} \cite{funt2010rehabilitation} \cite{joze2012role} for illumination estimation. The illumination of each pixel is estimated as the mean or maximum value of all pixels within the neighborhood. Besides, LIME in \cite{guo2017lime} estimates the illumination of each pixel as the maximum value of R, G and B channels.

Learning-based methods aim to learn the regularities of illumination and reflectance from training data and leverage them to improve the generalization of Retinex algorithms. For example, L Shen \emph{et al. } \cite{Shen2017} proposes MSR-Net by combining the MSR \cite{jobson1997multiscale} with the feedforward convolution neural network. S Park \emph{et al.} \cite{park2018dual} constructs a dual autoencoder network based on the Retinex theory to learn the regularities of illumination and noise respectively. A deep Retinex-Net is proposed in \cite{wei2018deep} to learn the key constraints including the smoothness of illumination, and the consistent reflectance shared by paired low/normal-light images. KG Lore \emph{et al.} \cite{lore2017llnet} proposes a LLNet to learn the contrast-enhancement and denoising relevant regularities simultaneously.

Image noise inevitably exists in low-light images, which will be amplified after contrast enhancement. L Li \emph{et al.} \cite{li2015low} improves the visual quality further by segmenting the observed image into superpixels and adaptively denoising different segments via BM3D \cite{dabov2007image}. After removing illumination effects, BM3D \cite{dabov2007image} is executed in \cite{guo2017lime}\cite{Shen2017} to suppress the amplified noise on dark regions. Besides, M Elad \emph{et al. } \cite{elad2005retinex} involves two bilateral filters on the modified Retinex model and transfers the illumination estimation and denoising into a progressive programming problem. However, the coupling relation between illumination and noise is neglect in these methods, which will lead to inaccurate estimation of illumination or noise, resulting in under or over-smoothing
in the enhancement results.

Different from the existing methods, we propose a progressive framework to perceive the illumination and noise of low-light image in a mutually reinforced manner. Particularly, the perception models are achieved by fully pointwise convolutional units, which extract  the representative pixels corresponding to illumination or noise from the input images. This modeling strategy can be interpreted as building unnatural representations for low-light images, which not only suppress the interference caused by ambiguity between tiny textures and image noises, but also improves the computational efficiency.

\section{Proposed Method}

\subsection{Motivation}
Referring to \cite{elad2005retinex}, low-light image enhancement task can be described as the following optimization problem:
\begin{equation}
\small
\begin{array}{l}
\mathop {\min }\limits_{\ell ,r:\ell  \ge s} \left\{ {{\lambda _\ell }\left\| {\ell  - s} \right\|_2^2 + \rho (\ell )} \right\} + \alpha \left\{ {{\lambda _r}\left\| {r - s + \ell } \right\|_2^2 + \rho (r)} \right\},
\end{array}
\label{eq:TwoBilateralFramwork}
\end{equation}
where $s$, $l$, and $r$ are low-light image, illumination and reflectance, $\alpha$ is a weight parameter. One feasible solution is to decompose the optimization into two sub-tasks, \emph{i.e.}, illumination estimation and noise level estimation, which can be optimized alternatively as follows:

\begin{equation}
\begin{array}{l}
\mathop {\min }\limits_{\ell :\ell  \ge s} \left\{ {{\lambda _\ell }\left\| {\ell  - s} \right\|_2^2 + \rho (\ell )} \right\} ,
\end{array}
\label{eq:IlluminationEstimation-FirstStage}
\end{equation}

\begin{equation}
\begin{array}{l}
\mathop {\min }\limits_{r:\ell  \ge s} \alpha \{ {\lambda _r}\left\| {r - s + \ell } \right\|_2^2 + \rho (r)\} ,
\end{array}
\label{eq:IlluminationEstimation-SecondtStage}
\end{equation}
where $\rho ( \cdot )$ serves as a quadratic form regularized constraint which represents the local statistical properties of ambient illumination and image noise. Thus, modeling statistical regularities of an input low-light image plays an essential role in Retinex based image enhancement.

Modeling statistical regularities for low-light image enhancement can be interpreted as to select the most representative pixels for illumination and noise from input images, and thus to extract an unnatural representation for them. For low-light images, the representative pixels for illumination usually refer to the regions with high reflectance in the scene, which includes white (grey) or specularity area, \emph{e.g.} sky, window and road surfaces, and thus have high intensity on all the three color channels \cite{Zhang_2017_CVPR}. Moreover, the noises existing in low-light images are mainly caused by dark current, electronics shot and photon in camera imaging \cite{liu2008automatic}\cite{healey1994radiometric}, and are colored noises whose intensity distributions depend on ambient illumination. Therefore, the representative pixels for low-light noises should have different intensity distribution across three color channels from their neighborhoods.

Based on the above observation, we find that the intensity distribution across color channels can be leveraged for the extraction of representative pixels from low-light images. Therefore, we devise a novel CNN model consisting of pointwise convolutional units to represent the statistics for illumination and noise of low-light image. The principle of a pointwise convolutional layer is illustrated in Fig. \ref{fig:Motivation-1}. The inherent correlations among the three color components of each pixel are firstly represented by pointwise convolution, and the statistically significant pixels are then selected in the representation space by Max pooling operation. The combined effect of multiple pointwise convolution layers is equivalent to multi-representation fusion and non-maximum suppression in different receptive fields. The statistics of the final selected representative pixels are modeled as regularity for Retinex decomposition.

\begin{figure}
\begin{center}
\includegraphics[width=0.945\linewidth]{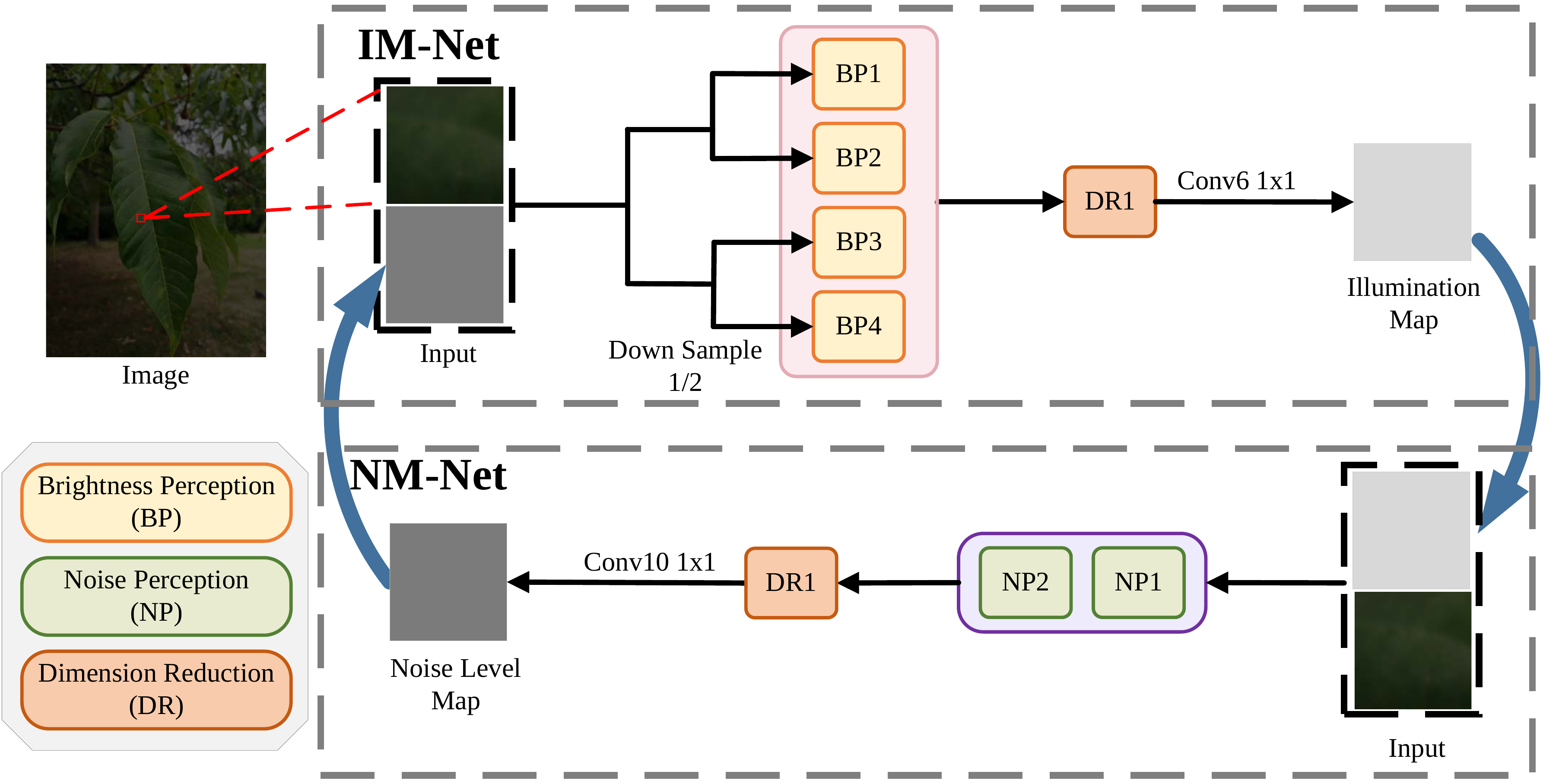}
\end{center}
   \caption{An overview of the proposed illumination-noise perception network for progressive Retinex framework.}
\label{fig:fig2-iterativeFramework}
\end{figure}

Furthermore, due to the coupling relation between illumination and noise, there inevitably exists an intersection between the selected representative pixels for illumination and noise, which may introduce the interference for statistical modeling. To address this issue, we present a progressive learning mechanism, in which the representative pixels for illumination and noise are selected and statistically modeled in a mutually reinforced manner. The illumination map can be used as guidance to suppress the interference of intersection pixels for noise level estimation, and vice versa. As shown in Fig. \ref{fig:fig2-iterativeFramework}, the two processes are progressive performed until getting stable results. Each of the preceding modeling processes benefits from the gradual improvement results in the other.

\label{subsec:iterativeFramework}

\subsection{Network Architecture}
The proposed progressive framework consists of two subnetworks: IM-Net and NM-Net, for illumination estimation and noise level estimation respectively. Besides, we present a fully pointwise convolutional unit to select the representative pixels related to illumination or noise, and thus model the statistics of these pixels as the regularity $\rho (*)$ in Eq.(3) and (4).  Note that our method is quite different with the point-wise CNN in \cite{zhang2019famed}\cite{zhang2018fully}, which  includes a pixel scrambling step and aims at achieving a lightweight CNN.

\textbf{Pointwise convolution\quad}
By performing convolution on RGB channels, the pointwise kernel can map all pixels to the same representation space. Without introducing spatial structures, the feature responses are only related to the pixel intensity on RGB channels\cite{zhang2019famed}\cite{zhang2018fully}. The pointwise kernel produces strong feature responses for the pixels within a specific intensity distribution. As a result, these pixels can be selected by non-maximum suppression after performing max pooling operation. A typical example of pixel selection process during illumination estimation is shown in Fig.~\ref{fig:Motivation-1}. As can be seen, every pointwise kernel can select a pixel set with similar intensity distribution. For example, as shown in Fig.~\ref{fig:Motivation-1}(a), the kernel ${K_{12}}$ mainly concentrates on the yellow pixels within the bottom left corner of the image, while ${K_{1N}}$ mainly selects the white pixels. By combining $N$ pointwise kernels together, the representative pixels for illumination can be selected completely, as shown in ${R_{1}}$ of Fig.~\ref{fig:Motivation-1}(a). The pixel confidence is proportional to the feature response. The pixels with lower confidence cannot accurately reflect the illuminant properties. These pixels can be further refined in the second layer as shown in Fig.~\ref{fig:Motivation-1}(c). With the network going deeper, the most representative pixels can be gradually selected. The statistical properties can be easily obtained from these pixels.

\textbf{The Structure of IM-Net\quad}The goal of IM-Net is to estimate the illumination for a given low-light patch. The pixels with maximum intensity are most related to ambient illumination. In order to select the most representative pixels, we present a fully pointwise CNN, \emph{i.e.}, IM-Net, to learn the representation of ambient illumination. We introduce two parallel convolutional branches as shown in Fig.~\ref{fig:fig2-iterativeFramework} to learn multi-scale features. A single branch consists of two parallel $1*1$ convolutional units to extract the statistics at each scale. Specifically, a $1*1$ convolutional unit contains a pointwise convolution process and a max pooling operation. Then, the features from the two branches are combined together as the input of the next $1*1$ convolutional unit. The details of IM-Net are shown in Table ~\ref{tab:IM-Net}. Based on the selected representative pixels, IM-Net can accurately model the statistical regularity of ambient illumination and estimate the illumination.

\begin{table}[t]
  \centering
  \caption{The details of each component of IM-Net.}
  \small
    \begin{tabular}{llllll}
    \hline
          & Input Size  & Num   & Filter & Stride & pad \\
    \hline
    Conv-BP1 & 3x32x32 & 160   & 1x1   & 1     & 0 \\
    MaxPool-BP1 & 160x32x32 & -     & 16x16 & 16    & 0 \\
    Conv-BP2 & 3x32x32 & 160   & 1x1   & 1     & 0 \\
    MaxPool-BP2 & 160x32x32 & -     & 20x20 & 16    & 2 \\
    Conv-BP3 & 3x16x16 & 160   & 1x1   & 1     & 0 \\
    MaxPool-BP3 & 160x16x16 & -     & 8x8   & 8     & 0 \\
    Conv-BP4 & 3x16x16 & 160   & 1x1   & 1     & 0 \\
    MaxPool-BP4 & 160x16x16 & -     & 10x10 & 8     & 1 \\
    Concat & 640x2x2 & -     & -     & -     & - \\
    Conv-DR1 & 640x2x2 & 80    & 1x1   & 1     & 0 \\
    MaxPool-DR1 & 80x2x2 & -     & 2x2   & 2     & 0 \\
    Conv6 & 80x1x1 & 1     & 1x1   & 1     & 0 \\
    \hline
    \end{tabular}%
  \label{tab:IM-Net}%
\end{table}%

\begin{table}[t]
  \centering
  \caption{The details of each component of NM-Net.}
    \small
    \begin{tabular}{llllll}
    \hline
          & Input Size  & Num   & Filter & Stride & pad \\
    \hline
    Conv-NP1 & 3x32x32 & 160   & 1x1   & 1     & 0 \\
    MaxPool-NP1 & 160x32x32 & -     & 4x4   & 4     & 0 \\
    Conv-NP2 & 160x8x8 & 160   & 1x1   & 1     & 0 \\
    MaxPool-NP2 & 160x8x8 & -     & 4x4   & 4     & 0 \\
    Conv-DR2 & 160x2x2 & 80    & 1x1   & 1     & 0 \\
    MaxPool-DR2 & 80x2x2 & -     & 2x2   & 2     & 0 \\
    Conv10 & 80x1x1 & 1     & 1x1   & 1     & 0 \\
    \hline
    \end{tabular}%
  \label{tab:NM-Net}%
\end{table}%

\textbf{The Structure of NM-Net\quad}The goal of NM-Net is to estimate the noise variance for a given low-light patch. The variance on local patch can be formulated as:
\begin{equation}
{\sigma ^2} = \sigma _n^2 + {\sigma _I^2} ,
\label{eq:NoiseVariance}
\end{equation}
where $\sigma ^2$ is the variance of all pixels within local patch, $\sigma _n^2$ is the noise variance and $\sigma _I^2$ is the texture variance.  Conventional algorithms often treat all pixels equally and use the variance of all pixels within local patch to represent the noise variance. But it often leads to an overestimation of real noise level. In order to select out the most representative pixels for noise level estimation, we propose a fully pointwise CNN, \emph{i.e.}, NM-Net. The details of NM-Net are shown in Table ~\ref{tab:NM-Net}. Based on the selected representative pixels, NM-Net can accurately model the statistical regularity of noise.

\textbf{Progressive Mechanism\quad}
Because of the coupling relation between illumination and noise, there inevitably exists an intersection between the selected representative pixels for illumination and noise. In order to suppress the mutual effect between illumination and noise, we propose a progressive mechanism to perceive the inherent relation between illumination and noise. As shown in Fig.~\ref{fig:fig2-iterativeFramework}, the IM-Net is used as the first stage for illumination estimation. The input of the IM-Net in the current iteration is the estimated noise level from previous iteration, together with the original patch. Specifically, the noise level input of IM-Net in the first iteration is set to 0 manually. The noise level map has unified statistical property with noise relevant pixels, which can be used as a reference to suppress the response of noise relevant pixels in the first stage. Thus, the IM-Net can learn better representation for illumination distribution. After the first stage completed, the estimated illumination map is transmitted to the second stage, \emph{i.e.}, NM-Net, together with the original patch. The estimated illumination map can also be used as the reference to guide the NM-Net to suppress the response of illumination relevant pixels. Each of the preceding two processes benefits from the gradual improvement results in the other. We progressively perform the two processes until obtaining stable results.

During the test phase, the input low-light image is going through the network which produces the illumination map and noise level map. They are resized to the original size same to the input using bilinear interpolation. To further smooth the illumination map, guided image filtering is used to suppress the artifacts \cite{he2013guided}. Besides, to quantitatively evaluate the accuracy of noise level estimation, we adopt BM3D \cite{dabov2007image} denoising algorithm, which has only one crucial parameter--noise level, but achieves SOTA performance. Under the guidance of the estimated noise level map, BM3D algorithm can separate the noise from the high-frequency details of input image.

\subsection{Loss Function}
We used mean squared error to supervise the network, which can be written as:
\begin{equation}
\begin{array}{l}
L = \frac{1}{n}\sum\limits_{i = 1}^n {{{\left\| {{X_i} - {Y_i}} \right\|}^2}} ,
\end{array}
\label{eq:LossFunction}
\end{equation}
where $n$ is the number of training samples, ${X_i}$ is the output of network and ${Y_i}$ is ground-truth.

\section{Experiments}

\subsection{Experimental Setup}
The IM-Net and NM-Net were trained for 50,000 iterations using Stochastic Gradient Descent with an initial learning rate $0.005$, weight decay $5e-6$ and momentum $0.9$, batch size of 128. The learning rate decreased by half from $0.005$ to $3.125e-4$ every $10000$ iterations. In IM-Net and NM-Net, the filter weights of each layer were initialized by MSRA \cite{he2015delving}. The IM-Net and NM-Net were implemented in Caffe \cite{jia2014caffe}.

\subsection{Synthetic Data Generation}
It is time-consuming and difficult to collect massive pairs of low-light and normal images of natural scenes (or the pairs of low-light images and their associated illumination and noise level maps). Instead, we resort to synthesized training data by simulating the low-light image generation process in cameras. Referring to \cite{liu2008automatic}, the noise distribution in low-light images can be modeled as:
\begin{equation}
\begin{array}{l}
I = f(L + {n_s} + {n_c}) ,
\end{array}
\label{NoiseDistribution}
\end{equation}
where $I$ is the captured image, $f( \cdot )$ is camera response function (CRF), $n_s$ accounts for the illumination-dependent noise, $n_c$ accounts for the independent noise. Besides, the real noise in low-light images is not white; however, there are spatial correlations introduced by ``demosaicing" \cite{ramanath2002demosaicking}. In order generate high fidelity low-light images, we propose a unified image generation model :
\begin{equation}
\begin{array}{l}
{I_l} = f(M({M^{ - 1}}({f^{ - 1}}(I_n\cdot t)) + {n_s} + {n_c}))) ,
\end{array}
\label{ImageGenerationModel}
\end{equation}
where $M(\cdot)$ and $M^{ - 1}(\cdot)$ denote Bayer pattern and inverse Bayer pattern respectively. $f^{ - 1}( \cdot )$ is the inverse camera response function.

\begin{figure}
\begin{center}
\includegraphics[width=0.9\linewidth]{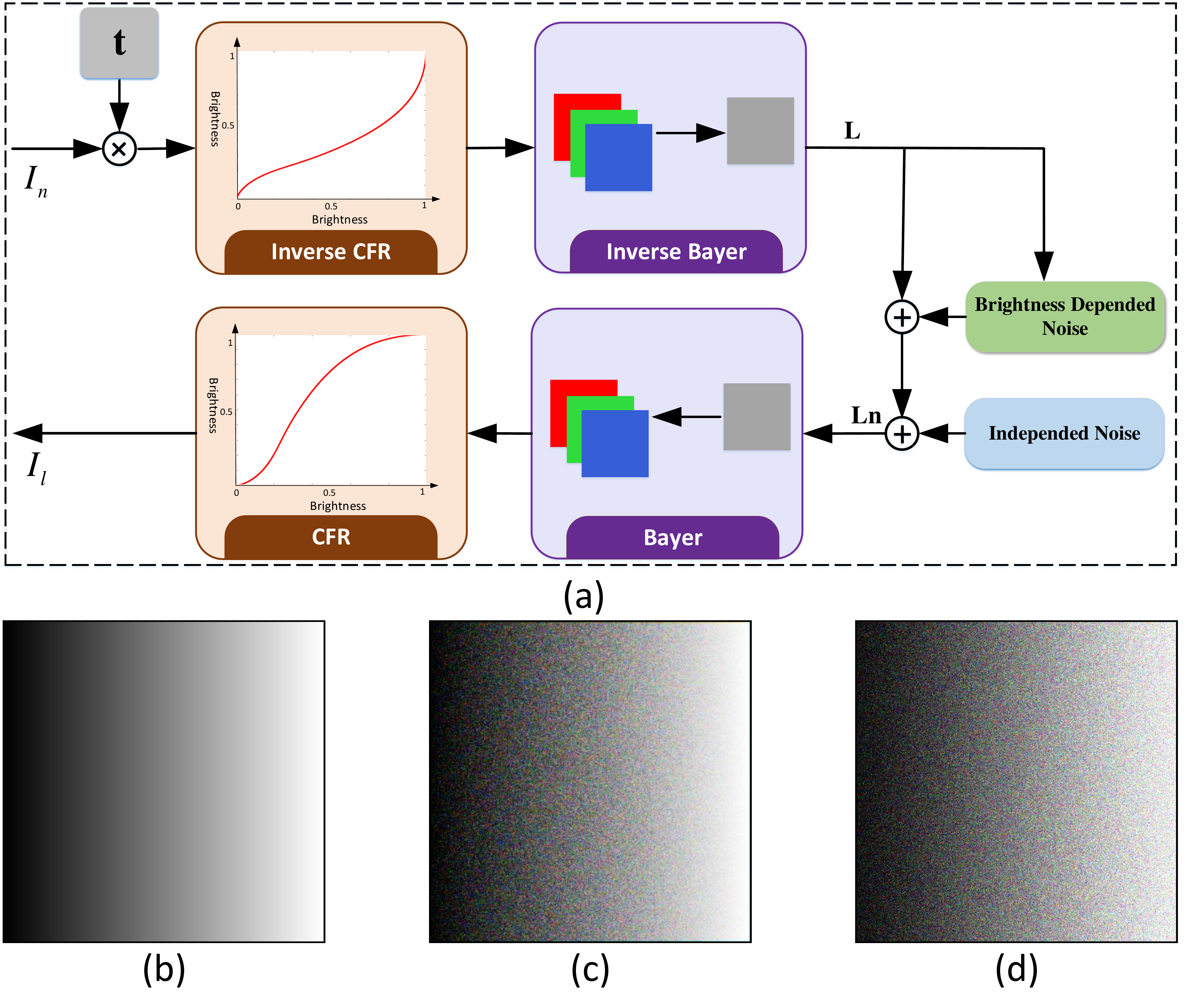}
\end{center}
   \caption{ (a) The image synthesis process described in Eq. (\ref{ImageGenerationModel}). (b) Test pattern. (c) Independent white noise synthetic result. (d) The synthetic result of (a).}
\label{fig:fig3-dataGeneration}
\end{figure}

The block diagrams of synthesizing low-light image is shown in Fig.~\ref{fig:fig3-dataGeneration}(a). Given a clear image patch $I_n$, the low-light condition is simulated by multiplying a coefficient $t$, which belongs to [0,1].  The irradiance $L$ is obtained after performing $f^{ - 1}( \cdot )$ and $M^{ - 1}(\cdot)$ on $I_n\cdot t$. Then, we successively add the illumination-dependent noise and independent noise to irradiance. After that, we perform $M(\cdot)$ and $f( \cdot )$ to obtain the synthesized low-light image patch. We collect $200$ well-exposed images from the Internet and extract 25000 image patches with size $32*32$. Grossberg, M.D \emph{et al}. \cite{grossberg2004modeling} provides $201$ kinds of CRFs and we exploit the most widely used CRF50 and CRF60 in our experiments. Besides, the noise variances $\sigma _s$ and $\sigma _c$ are uniformly sampled from the ranges of [0; 0.16] and [0; 0.06], respectively. In this paper, we totally synthesize $250000$ low-light patches. We use $200000$ patches as the training set and $50000$ patches as the test set. The synthesized image, for the test pattern (Fig.~\ref{fig:fig3-dataGeneration}(b)), is shown in Fig.~\ref{fig:fig3-dataGeneration}(c). Compared with the synthetic result by using independent white noise in Fig.~\ref{fig:fig3-dataGeneration}(d), Fig.~\ref{fig:fig3-dataGeneration}(c) can better reflect the dependence between noise and illumination. 

\label{subsec:dataGeneration}

\subsection{Ablation Studies}
We evaluate the proposed method at different settings on $50$ synthetic images, \emph{i.e.}, the number of iterations between IM-net and NM-Net and different input. They are non-overlapping with the training and test set. Referring to \cite{cai2018learning}, we use the same 15-layer residual network as the base model.

\textbf{Number of Iterations\quad}To examine the improvements induced by different iterations, we experimentally compare models with different iterations on both synthetic images and real images. The peak signal-to-noise ratio (PSNR) and structural similarity (SSIM) \cite{wang2004image} between ground truth and enhanced results of models with different iterations are shown in Table~\ref{tab:addlabel}. Compared with IM-Net and NM-Net working independently, \emph{i.e.,} iteration is 0, the introduction of progressive mechanism can effectively improve the accuracy for illumination and noise level estimation. As the number of iterations increases, PSNR and SSIM gradually increase, and reach stability in the fourth iteration. This implies that the mutual effect between illumination and noise is progressively suppressed. The accuracy of the optimal iteration model is higher than that of the basic model, which demonstrates the effectiveness of the progressive mechanism for modeling the coupling relation between illumination and noise.

\begin{figure}
\begin{center}
\includegraphics[width=1\linewidth]{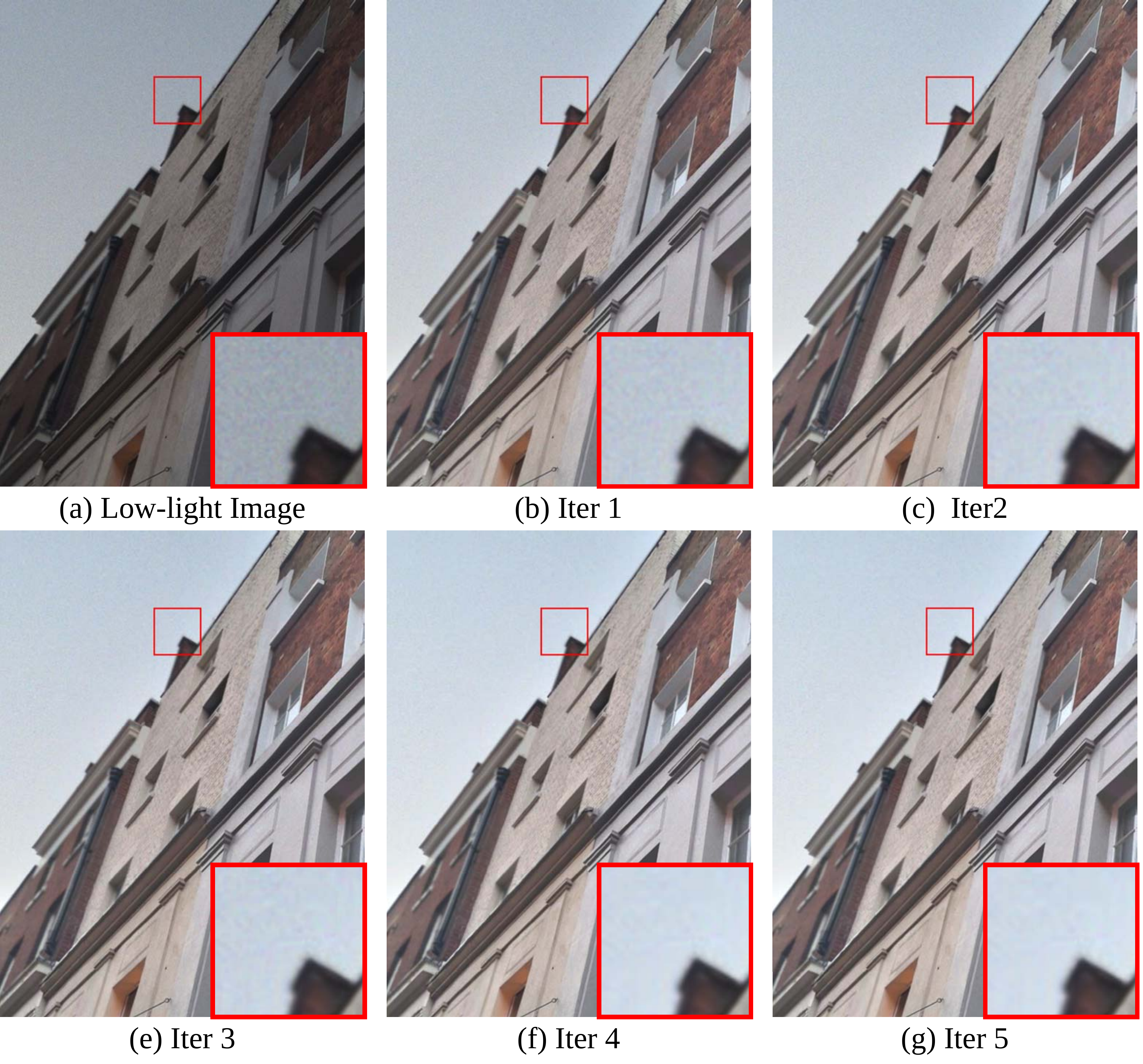}
\end{center}
   \caption{Enhanced images of models with different iterations on low-light image. (a) The original low-light image. (b)-(g) Enhanced images of our models with one iteration to five iterations.}
\label{fig:fig5-ORII}
\end{figure}

Figure \ref{fig:fig5-ORII} shows an example in real scene. As shown in Fig. \ref{fig:fig5-ORII}(b), the result generated by one iteration model is still low-light and contains noise. In contrast, these artifacts are gradually reduced as iterations increasing and it achieves stable on the fourth iteration, which proves the practicability of our method for real scenes. The model with four iterations is set as our default model in the following paper.

\label{subsec:ObjectiveOfRealImageOnIteration-1}

\begin{table}[t]
\small
  \centering
  \caption{The PSNR and SSIM of enhanced results on 50 synthetic low-light images of models with different iterations.}
    \begin{tabular}{cccccccc}
    \hline
          & Baseline & Iter 0 & Iter 1 & Iter 2 & Iter 3 & Iter 4 & Iter 5 \\
    \hline
    PSNR  & \textcolor[rgb]{0.00,0.00,1.00}{\textbf{22.85}} & 19.86 & 21.78 & 22.20 & 22.78 & \textcolor[rgb]{1.00,0.00,0.00}{\textbf{23.13}} & 23.12 \\
    \hline
    SSIM  & \textcolor[rgb]{0.00,0.00,1.00}{\textbf{0.907}} & 0.883 & 0.895 & 0.897 & 0.911 & \textcolor[rgb]{1.00,0.00,0.00}{\textbf{0.911}} & 0.910 \\
    \hline
    \end{tabular}%
  \label{tab:addlabel}%
\end{table}%

\begin{table}[t]
\small
  \centering
  \caption{The PSNR and SSIM of enhanced results by using intermediate enhancements as network input.}
  \setlength{\tabcolsep}{1.9mm}{
    \begin{tabular}{cccccccc}
    \hline
     Model    & Iter 1 & Iter 2 & Iter 3 & Iter 4 & Iter 5 & Proposed\\
    \hline
    PSNR  & 20.69 & 17.56 & 16.46 & 15.87 & 15.50 & \textcolor[rgb]{1.00,0.00,0.00}{\textbf{23.13}}\\
    \hline
    SSIM  & 0.797 & 0.761 & 0.740 & 0.725 & 0.714 & \textcolor[rgb]{1.00,0.00,0.00}{\textbf{0.9111}}\\
    \hline
    \end{tabular}
  \label{tab:addlabe1-1}}
\end{table}

\begin{table}[t]
\small
  \centering
  \caption{The PSNR and SSIM of enhanced results by using constant noise level to substitute the NM-Net.}
  \setlength{\tabcolsep}{0.8mm}{
    \begin{tabular}{cccccccc}
    \hline
     Noise Level     & 0 & 0.04 & 0.08 & 0.12 & 0.16 & 0.20 & Proposed\\
    \hline
    PSNR  & 18.32 & 20.84 & 21.66 & \textcolor[rgb]{0.00,0.00,1.00}{\textbf{22.55}} & 22.07 & 21.6 & \textcolor[rgb]{1.00,0.00,0.00}{\textbf{23.13}}\\
    \hline
    SSIM  & 0.760 & 0.770 & 0.825 & \textcolor[rgb]{0.00,0.00,1.00}{\textbf{0.868}} & 0.846 & 0.821 & \textcolor[rgb]{1.00,0.00,0.00}{\textbf{0.9111}} \\
    \hline
    \end{tabular}%
  \label{tab:addlabe1-2}}%
\end{table}%

\textbf{Network Input\quad}We perform an experiment on 50 synthetic images by using the intermediate enhancements, \emph{i.e.}, illumination enhancement or denoising result in every iteration. The results are shown in Table \ref{tab:addlabe1-1}. Due to the accumulation error, the testing results are getting worse with the increase of iteration times, which is inferior to our proposed model. Thus, we use the estimated noise level (or illumination) from previous iteration, coupling with the original patch as the input of the IM-Net (or NM-Net). And we only perform the image enhancement and denoising at the last iteration.

\begin{table*}[t]
  \centering
  \normalsize
  \setlength{\tabcolsep}{1.8mm}{
  \caption{The NIQE results on IP100 dataset and the PSNR/SSIM results on FNF38 dataset.}
    \begin{tabular}{cccccccccc}
    \hline
          &       & SRIE   & JIEP & LIME  & HQEC  & SRLL  & NPE  & Baseline  & Ours \\
          \hline
    \multirow{2}[0]{*}{FNF38} & PSNR  & \textcolor[rgb]{0.00,0.00,1.00}{\textbf{19.98}} & 19.64 & 16.89 & 18.89 & 19.35 & 17.71 & 19.68 & \textcolor[rgb]{1.00,0.00,0.00}{\textbf{20.12}} \\
    \cline{2-10}
          & SSIM  & \textcolor[rgb]{0.00,0.00,1.00}{\textbf{0.81}}  & 0.80   & 0.78  & 0.79  & 0.80   & 0.78  & 0.80  & \textcolor[rgb]{1.00,0.00,0.00}{\textbf{0.82}} \\
          \hline
    IP100 & NIQE  & 3.66  & 3.58  & 4.20   & 3.85  & 3.45  & 4.42  & \textcolor[rgb]{0.00,0.00,1.00}{\textbf{3.44}}  & \textcolor[rgb]{1.00,0.00,0.00}{\textbf{3.34}} \\
    \hline
    \end{tabular}%
  \label{tab:NIQEPSNRSSIM}}%
\end{table*}%

\begin{table}[htbp]
  \scriptsize
  \centering
  \caption{The PSNR/SSIM results on LOL dataset and the NIQE results on NPE dataset.}
    \begin{tabular}{ccccccccc}
    \hline
          &       & SRIE  &   JIEP & LIME  & HQEC  &   SRLL &   NPE & Ours \\
    \hline
    \multirow{2}[0]{*}{LOL   } & PSNR  & 11.8552 & 12.0466 & \textcolor[rgb]{0.00,0.00,1.00}{\textbf{17.1818}} & 16.6241 & 13.8765 & 16.6972 & \textcolor[rgb]{1.00,0.00,0.00}{\textbf{18.8025}} \\
    \cline{2-9}
          & SSIM  & 0.4979 & 0.5124 & 0.6349 & 0.6079 & \textcolor[rgb]{0.00,0.00,1.00}{\textbf{0.6577}} & 0.5945 & \textcolor[rgb]{1.00,0.00,0.00}{\textbf{0.7215}} \\
    \hline
    NPE   &  NIQE & \textcolor[rgb]{0.00,0.00,1.00}{\textbf{2.8912}} & 2.9174 & 3.2606 & 3.075 & 3.8476 & 2.8955 & \textcolor[rgb]{1.00,0.00,0.00}{\textbf{2.7519}} \\
    \hline
    \end{tabular}%
  \label{tab:NPELOL}%
\end{table}%

\label{subsec:ObjectiveComparison1}
\begin{figure*}
\begin{center}
\includegraphics[width=0.9\linewidth]{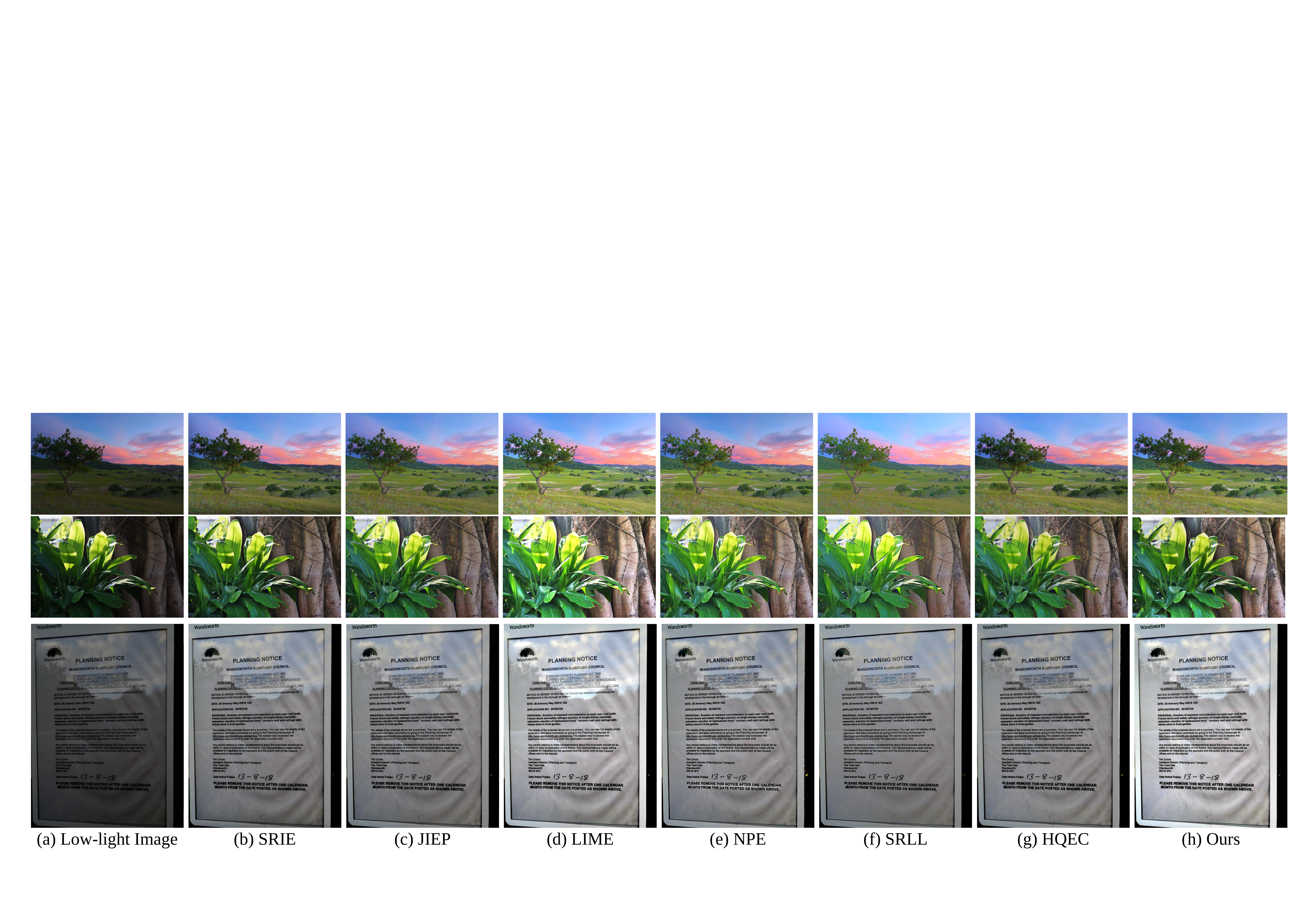}
\end{center}
   \caption{The contrast enhancement results of state-of-the-art methods including SRIE \cite{fu2016weighted}, JIEP  \cite{cai2017joint}, LIME \cite{guo2017lime}, NPE \cite{wang2018naturalness}, SRLL \cite{li2018structure}, HQEC \cite{zhang2018high}, and the proposed method.}
\label{fig:fig6-OCR1}
\end{figure*}

\textbf{Noise Level\quad}We also perform an experiment on 50 synthetic images by using constant noise level to replace the NM-Net. Table \ref{tab:addlabe1-2} shows the experiment results. Compared with using IM-Net only, \emph{i.e.}, the noise level is 0, the quantitative results are significantly improved after denoising. However, the denoising results guided by the constant noise level are inferior to the result guided by the NM-Net, which demonstrates the significance of the noise level estimation network.

\subsection{Comparisons with SOTA Methods}
To verify the superiority of our method, we compare it with the SOTA methods, including SRIE \cite{fu2016weighted}, LIME \cite{guo2017lime}, JIEP \cite{cai2017joint}, HQEC \cite{zhang2018high}, SRLL \cite{li2018structure}, NPE \cite{wang2018naturalness} and LDSE \cite{cai2018learning}. We perform the experiments on two collected datasets (IP100 and FNF38) and three public datasets (MPI \cite{cai2018learning}, LOL \cite{wei2018deep}, NPE \cite{wang2018naturalness}).

The IP100 dataset consists of two parts: ICI35 and P-65. ICI35 contains 35 identified challenging low-light images collected from previous works \cite{cai2017joint}. P-65 includes the other 65 challenging images which are captured by Huawei P20 smartphones. The ground-truth of images in IP100 are unavailable, and thus we adopt the widely used blind image quality assessment, \emph{i.e.}, natural image quality evaluator (NIQE) \cite{hautiere2011blind}, to evaluate the enhanced results. The lower NIQE value means higher image quality. FNF38 dataset includes 38 ambient and flash illumination pairs, which are selected from the FAID dataset \cite{aksoy2018dataset}. The two images from the same pair are well-aligned and the ambient image can be used as the ground truth. So we use the PSNR and SSIM to assess the enhanced result for this dataset. MIP dataset \cite{cai2018learning} contains 589 high-resolution multi-exposure sequences with 4,413 images. And the ground truth of this dataset is derived from several representative multi-exposure image fusion and stack-based high dynamic range imaging algorithms \cite{raman2009bilateral}\cite{shen2011generalized}\cite{zhang2012gradient}\cite{shen2014exposure}\cite{kou2017multi}\cite{bruce2014expoblend}.

\textbf{Objective Comparisons\quad}Table ~\ref{tab:NIQEPSNRSSIM} lists the NIQE and PSNR/SSIM results of different methods on IP100 and FNF38 datasets. As can be seen, the proposed method achieves lower NIQE and higher PSNR/SSIM than all of the other methods on two datasets. This implies that our method has better generalization for various scenes than other prior/assumption based methods. Next, we compare our method with CNN-based single image contrast enhancer \cite{cai2018learning} on MPI dataset. We achieve a slightly better result (PSNR: 19.77, FSIM \cite{zhang2011fsim}: 0.9456) than the method in \cite{cai2018learning} (PSNR: 19.77, FSIM: 0.9347) for under-exposure image enhancement. It demonstrates the superiority of our progressive network over the existing architectures. We also evaluate the performance of our method on LOL and NPE datasets, and the contrastive experimental results are listed in Table ~\ref{tab:NPELOL}. Our method demonstrates the best performances in all three terms.

\label{subsec:ObjectiveComparison2}
\begin{figure*}
\begin{center}
\includegraphics[width=0.88\linewidth]{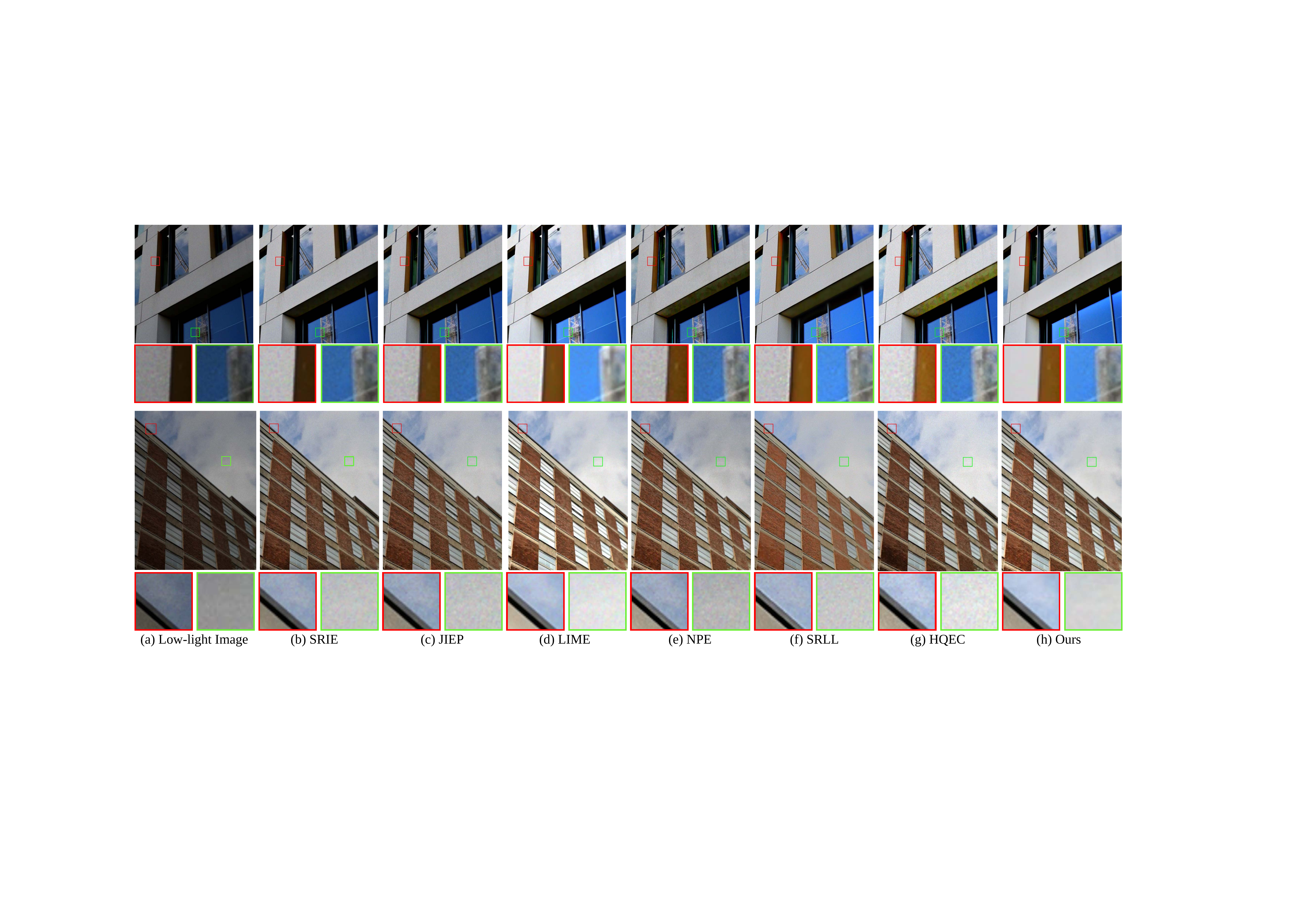}
\end{center}
   \caption{The contrast enhancement and denoising results of state-of-the-art methods including SRIE \cite{fu2016weighted}, JIEP  \cite{cai2017joint}, LIME \cite{guo2017lime}, NPE \cite{wang2018naturalness}, SRLL \cite{li2018structure}, HQEC \cite{zhang2018high}, and the proposed method.}
\label{fig:fig7-OCR2}
\end{figure*}

\begin{figure}
\begin{center}
\includegraphics[width=0.85\linewidth]{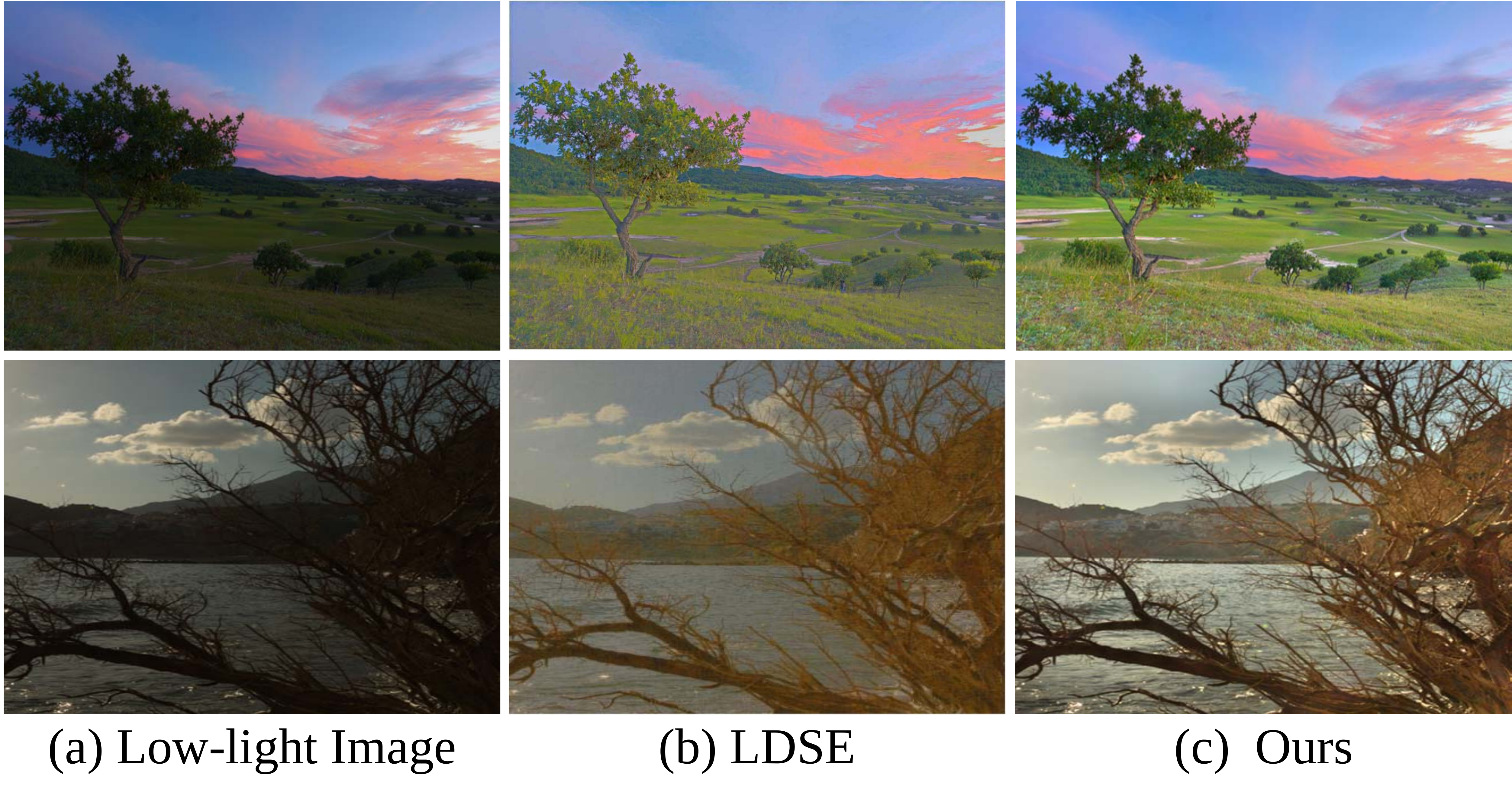}
\end{center}
   \caption{The contrastive experiment with LDSE \cite{cai2018learning}. (a) Input low-light image. (b) Results of LDSE \cite{cai2018learning}. (c) Results of the proposed method.}
\label{fig:fig8-LeiZhang}
\end{figure}

\textbf{Subjective Comparisons\quad}Figure.~\ref{fig:fig6-OCR1} shows the visual results of low-light images with little noise. We mainly concentrate on the contrast enhancement on these images. We can see that although methods such as SRIE, JIEP, NPE and HQEC can improve the contrast to a certain extent, the enhancement results still have a certain degree of low-light. The LIME algorithm can efficiently remove the unfavorable illumination and improve the global contrast. However, the enhanced result tends to exhibit over-exposure in some regions. SRLL can improve the global contrast but will introduce color cast. We also compare with LDSE on MIP dataset in Fig.~\ref{fig:fig8-LeiZhang}. The LDSE can improve the overall visibility of scenes but introduces the color cast and blur. Compared with other methods, our method can efficiently remove the unfavorable illumination without introducing over-exposure or color cast. It demonstrates the proposed method has superior capacity for illuminant statistical modeling.

\label{subsec:visio-LeiZhang}

Figure. ~\ref{fig:fig7-OCR2} shows the results on noisy low-light images captured in real scenes. Previous methods, like HQEC and NPE, are mainly designed for the contrast enhancement and do not include the specific operations for noise removal. With contrast improving, the noise is amplified in the enhanced result. The LIME algorithm uses BM3D to suppress the amplified noise in the enhancement result, but there is still some noise in the result. This is because globally consistent noise level coefficients cannot be applied to all regions of an image. The SRLL algorithm introduces the noise term into a variational Retinex model to formulate the captured low-light and noise images. This method can only suppress the noise to a certain degree and the enhanced result still contains the residual noise. Compared with other methods, our method can remove noise adequately while enhancing contrast, without introducing the over- or under-denoising problems. It demonstrates the superior capacity of our method for modeling the coupling relation between illumination and noise level.

\textbf{Computational Efficiency\quad} Moreover, we also compare the computational efficiency with the CNN based method \cite{cai2018learning}. Our progressive framework is found to be 50 faster than \cite{cai2018learning}. It processes a 129*129*3 image in 0.46s (run on CPU), compared to 26.47s in \cite{cai2018learning}. Besides, the processing time can be further reduced to 0.03s with GPU acceleration. This advantage is due to the fully pointwise convolutional structure, which can be implemented efficiently.

\section{Conclusion}
In this paper, we present a progressive Retinex framework to improve the quality of low-light image in a mutually reinforced manner. The proposed framework is implemented based on fully pointwise convolutions, which can suppress the interferences caused by ambiguity between tiny textures and image noises, and achieve high computational efficiency. The comprehensive evaluations on synthetic and real low-light images demonstrate that our proposed method achieves superior performance over the representative state-of-the-art low-light image enhancement methods.

The limitation of our proposed method is that it only captures the statistical distribution in the pixel space while neglects the structural properties. One feasible solution to this issue is to design a multi-branch network, which is able to perceive both the inherent structural and statistical properties of low-light images together. We leave it as our future work.

\section*{Acknowledgments}
\small
\noindent{This work was supported by the National Key R$\&$D Program of China under Grant 2017YFB1300201, the National Natural Science Foundation of China (NSFC) under Grants 61622211, 61620106009, 61872327, 61472380 and 61806062 as well as the Fundamental Research Funds for the Central Universities under Grant WK2100100030 and WK2380000001.}


\clearpage
%
\bibliographystyle{ACM-Reference-Format}


%
\appendix


%
%
%
%
%
%

\end{document}